\documentstyle[12pt,aaspp4,flushrt]{article}

\catcode`\@=11 
\def\@versim#1#2{\vcenter{\offinterlineskip
        \ialign{$\m@th#1\hfil##\hfil$\crcr#2\crcr\sim\crcr } }}
\newcommand{\beq}{\begin{equation}}
\newcommand{\eeq}{\end{equation}}
\def\lsim{\mathrel{\mathpalette\@versim<}}
\def\gsim{\mathrel{\mathpalette\@versim>}}

\begin{document}
\title{A Thermal Bremsstrahlung Model For the Quiescent X-ray Emission from
Sagittarius A*} \author{Eliot Quataert} \affil{
UC Berkeley, Astronomy Department, 601 Campbell Hall, Berkeley, CA
94720; eliot@astron.berkeley.edu}

\medskip

\begin{abstract}

I consider the thermal bremsstrahlung emission from hot accretion
flows (Bondi/ADAFs), taking into account the finite size of the
observing telescope's beam ($R_{beam}$) relative to the Bondi
accretion radius ($R_{A}$). For $R_{beam} \gg R_{A}$ soft X-ray
emission from the hot interstellar medium surrounding the black hole
dominates the observed emission while for $R_{beam} \ll R_{A}$ hard
X-ray emission from the accretion flow dominates. I apply these models
to {\it Chandra} observations of the Galactic Center, for which
$R_{beam} \approx R_{A}$.  I argue that bremsstrahlung emission
accounts for most of the ``quiescent'' (non-flaring) flux observed by
{\it Chandra} from Sgr A*; this emission is spatially extended on
scales $\sim R_{A} \sim 1''$ and has a relatively soft spectrum, as is
observed.  If accretion onto the central black hole proceeds via a
Bondi or ADAF flow, a hard X-ray power law should be present in deeper
observations with a flux $\sim 1/3 $ of the soft X-ray flux;
nondetection of this hard X-ray component would argue against
ADAF/Bondi models. I briefly discuss the application of these results
to other low-luminosity AGN.

\

\noindent {\it Subject Headings:} Galaxy: center --- accretion,
accretion disks

\end{abstract}

\section{Introduction}

Spherical Bondi accretion predicts that the interstellar medium around
a black hole should be gravitationally captured on scales of $R_A
\approx GM/c^2_s$, where $M$ is the mass of the black hole and $c_s$
is the sound speed of gas in the vicinity of $R_A$ (e.g., Bondi \&
Hoyle 1944; Shvartsman 1971).  The same is true for hot accretion flow
models that include dynamically important angular momentum, such as
advection-dominated accretion flows (ADAFs; e.g., Ichimaru 1977; Rees
et al. 1982; Narayan \& Yi 1994) and its variants (e.g., Blandford \&
Begelman 1999).  For radii $\lsim R_A$ the captured gas accretes onto
the central black hole and the dynamics is determined by accretion
physics rather than interstellar medium physics.

At a minimum, the gas in Bondi and ADAF models emits thermal
bremsstrahlung emission from radii $\lsim R_A$ (e.g., Narayan et
al. 1998; Di Matteo et al. 1999, 2000).  Additional X-ray emission
from synchrotron or inverse Compton processes in the accretion flow or
jet can also be present depending on the accretion rate and the
efficiency of electron acceleration (e.g., Narayan et al. 1998;
Markoff et al. 2001; Yuan et al. 2002; Narayan 2002).  For
sufficiently low-luminosity systems bremsstrahlung emission may
dominate over other emission processes in the X-ray band; e.g.,
Narayan et al. (1999; see Fig. 6) found that bremsstrahlung dominated
for $L_X \lsim 10^{-8} L_{\rm Edd}$.  This range of luminosities is
now routinely probed by {\it Chandra} observations (e.g., Ho et
al. 2001).  In particular, for the Galactic Center $L_X \sim 10^{-11}
L_{\rm Edd}$ (Baganoff et al. 2002) and so it is {\it a priori}
plausible that bremsstrahlung contributes significantly to the
observed emission.

For a given system, the relative contribution of the accretion flow
($R \lsim R_A$) and the ambient medium ($R \gsim R_A$) to the thermal
bremsstrahlung emission depends on the size of the observing
telescope's beam in units of the Bondi accretion radius.  For
$R_{beam} \gg R_A$ the ambient medium dominates the observed emission
while for $R_{beam} \ll R_A$ the accretion flow does.  The
interpretation of observed data therefore depends sensitively on
$R_{beam}/R_A$.

For ambient temperatures of $\approx 1$ keV, $R_A \approx 0.07$ pc for
the $2.6 \times 10^6 M_\odot$ black hole at the Galactic Center and
$\sim 30$ pc for the $\sim 10^9 M_\odot$ black holes in massive
elliptical galaxies such as M87, NGC 4472, and NGC 1399 (for the
Galactic Center, see Genzel et al.  1997 and Ghez et al. 1998 for a
black hole mass estimate and Baganoff et al. 2002 for a temperature
measurement; for elliptical galaxies, see, e.g., Gebhardt et al. 2000
and Ferrarese \& Merritt 2000 for black hole mass estimates and
Loewenstein et al. 2001 for central temperature measurements).  For
comparison, the $\approx 1"$ angular resolution of the {\it Chandra}
X-ray Observatory corresponds to a distance of $\approx 0.04$ pc at
the Galactic Center and $\approx 85$ pc in nearby X-ray clusters such
as Virgo and Fornax.  {\it Chandra} observations thus probe length
scales comparable to the Bondi accretion radius for a number of
supermassive black holes.  In this paper I present models for the
bremsstrahlung emission from hot accretion flows that can be applied
to {\it Chandra} observations with $R_{beam} \sim R_A$.

This paper is organized as follows.  In the next section (\S2) I show
X-ray spectra for bremsstrahlung emission from Bondi accretion flows
that explicitly account for the finite size of the observing
telescope's beam.  I then compare these predictions with {\it Chandra}
observations of Sgr A* at the Galactic Center (\S3). In \S4 I conclude
and discuss additional applications of these results.

\section{Predicted Spectra}

Figure 1 shows the density and temperature profiles for spherical
accretion onto a central black hole; the radius is in units of the
Bondi accretion radius, the temperature is in Kelvin, and the density
units are arbitrary. In these models the accretion flow is
self-consistently ``matched'' onto an external ambient medium; this is
important when considering observations with $R_{beam} \sim R_A$ since
one cannot always assume that the asymptotic ($R \ll R_A$) accretion
flow structure is directly observed.  The solid lines in Figure 1 are
for the original Bondi problem of accretion from a uniform medium,
while the dotted lines are for accretion from a stratified medium in
which the ambient density decreases with distance from the central
black hole as $\rho \propto R^{-1}$ for $R \gsim R_A$ (these results
are based on calculations described in Quataert \& Narayan 2000).  For
both models in Figure 1 the ambient medium is assumed to have a
temperature of $\approx 1$ keV.  Although it needs to be checked on a
case by case basis, the stratified models are probably a better
approximation of the real interstellar medium around supermassive
black holes.  For example, in X-ray clusters the density of gas
typically varies as $\sim R^{-1}$ (e.g., Fabian 1994), though {\it
Chandra} observations indicate that the density profile may flatten in
the central parts of some systems (see, e.g., Di Matteo et al. 2002
for M87).  For the Galactic Center I show below that the stratified
model is appropriate.

Although the detailed calculations presented in Figure 1 are for
spherical accretion without angular momentum, similar results are
expected for advection-dominated accretion flow models that include
angular momentum.  This is because ADAFs have density and temperature
profiles very similar to the Bondi models in Figure 1 (e.g., Narayan
\& Yi 1994).  Recent analytical and numerical work indicates, however,
that ADAF models may not describe the structure of radiatively
inefficient accretion flows (e.g., Blandford \& Begelman 1999;
Quataert \& Gruzinov 2000; Hawley \& Balbus 2002).  In particular the
density profiles in the above references are flatter than in Figure 1,
i.e., the density increases more slowly with decreasing radius for $R
\lsim R_A$. In \S4 I discuss the implications of these results for the
bremsstrahlung emission from hot accretion flows.

Figure 2 shows the X-ray bremsstrahlung spectra that result from the
density and temperature profiles in Figure 1; the left panel (dotted
lines) is for the stratified ambient medium while the right panel
(solid lines) is for the uniform ambient medium.  The spectra are
shown for different values of the size of the observing telescope's
beam in units of the Bondi accretion radius.  Each spectrum was
calculated using the standard non-relativistic bremsstrahlung formula
(e.g., Rybicki \& Lightman 1979).

For each spectrum in Figure 2 the density outside $R_{beam}$ was set
to zero so that the emission is dominated by regions along the line of
sight which are $\lsim R_{beam}$ from the central object.  In
practice, this requires that the density decrease at least as fast as
$R^{-1/2}$ far from the central object ($R \gsim R_{beam}$) so that
these regions do not contribute to the integral of the bremsstrahlung
emissivity along the line of sight. This is indeed the case for the
stratified models in Figure 2a since $\rho \propto R^{-1}$ at large
radii.  The uniform ambient medium models in Figure 2b are, however,
somewhat artificial because the spectra depend on the truncation of
the density outside $R_{beam}$.  They are included primarily to
illustrate the sensitivity of the predicted spectra to the structure
of the ambient medium on scales $\gsim R_A$.

Figure 2 shows how the bremsstrahlung spectra depend on
$R_{beam}/R_A$.  For $R_{beam} \gg R_A$ the ambient interstellar
medium dominates the observed emission and a soft X-ray spectrum is
expected (upper curves); the Bondi flow contributes a weak underlying
hard X-ray power law that can only be observed with a very high signal
to noise spectrum. For $R_{beam} \ll R_A$, on the other hand, the
accretion flow dominates the observed emission and the hard X-ray
spectrum characteristic of Bondi accretion flows is predicted (lower
curves).

A comparison of the results in Figures 2a \& 2b indicates that a
careful treatment of the radial stratification of the ambient medium
is important for quantitatively interpreting observational data: the
relative amount of soft X-ray and hard X-ray emission depends on both
$R_{beam}/R_A$ and the structure of the ambient medium on scales
$\gsim R_A$.


\section{Comparison with {\it Chandra} Observations of the Galactic Center}

As of January 2002, {\it Chandra} has observed the Galactic Center for
a total of $\approx 75$ ks over 2 epochs and has convincingly detected
a source coincident with the supermassive black hole (Baganoff et
al. 2001, 2002). For most of the 75 ks, Sgr A* was in a ``quiescent''
state characterized by a luminosity $\approx 2 \times 10^{33}$ ergs
s$^{-1}$ and a soft spectrum.  For a brief few ks period in the second
epoch, Sgr A* flared dramatically, reaching a luminosity of $\approx
10^{35}$ ergs s$^{-1}$ with a hard spectrum.

I focus here on the interpretation of the quiescent emission from the
Galactic Center, the properties of which can be summarized as follows:

\begin{enumerate}

\item{The source is {extended} with a diameter of $\approx 1"$.}

\item{The quiescent flux is the same in the first epoch and the second
epoch (separated by over a year) and before and after the flare.
Aside from the flare, and a possible precursor, the only evidence for
short timescale variability is $< 3 \sigma$.}

\item{Within 1.5" of Sgr A* the luminosity is $\approx 2 \times
10^{33}$ ergs s$^{-1}$ and the spectrum is quite soft, consistent with
a $2$ keV thermal plasma or a power law with a photon index of $\Gamma
= 1.5 - 2.7$ (see Fig. 2a).}

\item{Within 10" of Sgr A*, there is diffuse thermal bremsstrahlung
emission from hot gas surrounding the black hole (likely produced by
stellar winds; e.g., Coker \& Melia 1997).  The luminosity is $\approx
2 \times 10^{34}$ ergs s$^{-1}$ and the spectrum is consistent with a
$1.3$ keV thermal plasma, somewhat softer than the spectrum within
1.5".}

\end{enumerate}

As I now explain, bremsstrahlung emission provides a good description
of the quiescent {\it Chandra} observations of Sgr A* summarized in
items 1-4. Note that a 1.5" radius around Sgr A* corresponds to
$R_{beam} \approx R_A$, while a 10" radius corresponds to $R_{beam}
\approx 10 R_A$.

The ratio of the observed X-ray luminosities within 10'' and 1.5'' of
Sgr A* is only a factor of $\approx 10$.  Since bremsstrahlung
emission scales as $\sim R^3 \rho^2$, this implies that the typical
gas density at 10'' is $\approx 5$ times smaller than that at 1.5''.
Thus the density of hot gas around Sgr A* decreases roughly as $\sim
R^{-1}$ for $R \gsim R_A$ and the results in Figure 2a describe the
bremsstrahlung spectra in Bondi/ADAF models of Sgr
A*.\footnote{Without a radial density profile obtained by inverting
the observed surface brightness profile, this is the best that one can
do to estimate $\rho(R)$ at large radii.}  Because the gas density
decreases quite rapidly for $R \gsim R_A$ (faster than $R^{-1/2}$),
the emission observed by {\it Chandra} is dominated by gas within $R
\lsim R_{beam}$, as was assumed in Figure 2 (i.e., there is very
little contribution to the bremsstrahlung emissivity from gas at $R
\gsim R_{beam}$ along the line of sight).


In the temperature profiles shown in Figure 1 and used in Figure 2, I
fixed the temperature to be $\approx 1$ keV at large radii.  This
boundary condition was chosen so that the models are consistent with
{\it Chandra} observations of diffuse soft X-ray emission within 10''
of Sgr A* (item 4).  The density and luminosity units in Figures 1 and
2 are, however, arbitrary.  These results can therefore be applied to
any system for which the ambient temperature is $\sim 1$ keV.  For the
Galactic Center, a gas density of $\approx 20$ cm$^{-3}$ at 10''
reproduces the observed luminosity of $\approx 2 \times 10^{34}$ ergs
s$^{-1}$. The gas density at 1.5'' is then $\approx 100$ cm$^{-3}$.
With the temperature at large radii set by the 10'' {\it Chandra}
observations, and with the above normalization of the density at
several radii, there are no additional free parameters or boundary
conditions in Bondi accretion models. 

Figure 2a shows that the bremsstrahlung spectrum within 1.5'' of Sgr
A* is expected to be quite soft ($R_{beam} \approx R_A$).  It is,
however, somewhat harder than the spectrum within 10" ($R_{beam}
\approx 10 R_A$).  This is because the ambient gas is mildly
compressed and heated just inside the Bondi accretion radius. The soft
X-ray spectrum within 1.5'' of Sgr A*, and the slight increase in
temperature relative to the 10'' observations, are both in excellent
agreement with the {\it Chandra} observations (see items 3 \& 4 and
Fig. 2a).

Most existing predictions of the bremsstrahlung emission from
Bondi/ADAF models stress the hard X-ray power law seen at high
energies in Figure 2 (e.g., Narayan et al. 1998).  This hard X-ray
emission is produced at radii $\ll R_A$ where the flow structure
approaches its self-similar scalings of $\rho \propto R^{-3/2}$ and $T
\propto R^{-1}$.  For $R_{beam} \approx R_A$, this hard X-ray
component is a factor of $\sim 3$ times less luminous than the soft
X-ray emission produced by gas at $\sim R_A$ (Fig. 2a; middle curve).
Because of the limited photon statistics, the hard X-ray component
would be difficult to detect in present {\it Chandra} observations.
Instead, the observations are dominated by soft X-ray emission from
gas in the vicinity of the Bondi accretion radius.

As Yuan et al. (2002) noted, the bremsstrahlung interpretation
advocated here is strongly supported by the observation that the
quiescent X-ray source coincident with Sgr A* is resolved with a size
of $\approx 1" \approx R_A$ (item 1). Models that invoke synchrotron
or synchrotron self-Compton (SSC) emission for the majority of the
quiescent emission (e.g., Falcke \& Markoff 2000; Liu \& Melia 2001)
predict a source size $\sim 10-100$ Schwarzschild radii $\sim
10^{-4}-10^{-3}$ arcsec, much less than is observed. The
bremsstrahlung model also accounts naturally for the relatively
constant level of the quiescent flux (item 2) because the
characteristic variability timescale at $\sim R_A$ is $\sim 100$
years. 

Although bremsstrahlung emission probably dominates the quiescent flux
from Sgr A*, it cannot explain the dramatic X-ray flare observed from
the Galactic Center (Baganoff et al. 2001). Instead, this emission is
likely due to synchrotron or SSC from hot electrons very close to the
black hole (e.g., Markoff et al. 2001).  This suggests a model in
which bremsstrahlung from $\sim R_A$ provides the ``baseline'' flux
that dominates the quiescent emission, with a time variable
synchrotron and/or SSC contribution that is usually sub-dominant but
occasionally flares dramatically.  

\section{Discussion}

In this paper, I have presented model X-ray spectra for bremsstrahlung
emission from Bondi accretion flows that explicitly include the
contribution from hot ambient gas around the black hole
(Fig. 2). These spectra are useful for interpreting {\it Chandra}
observations of very low-luminosity AGN since theoretical models
suggest that bremsstrahlung may dominate over other emission processes
for $L \ll L_{Edd}$ (perhaps $L_X \lsim 10^{-8} L_{\rm Edd}$). Care
must be taken in interpreting observations of these systems because
the ambient gas around the black hole contributes significantly to the
observed emission if $R_{beam} \gsim R_A$ (see Fig. 2), as is the case
for most known systems even given {\it Chandra's} excellent angular
resolution.

I have applied these results to {\it Chandra} observations of Sgr A*
at the Galactic Center. I propose that, excluding the large X-ray
flare, bremsstrahlung emission from gas in the vicinity of the Bondi
accretion radius dominates the quiescent flux observed from Sgr
A*. This model explains why the observed spectrum is quite soft, why
the quiescent flux is relatively constant (i.e., independent of time),
and why the source is spatially extended (\S3). In this interpretation
the quiescent emission from Sgr A* does not presently constrain
accretion flow models because the emission arises from gas at $\sim
R_A$, i.e., from the ``transition region'' between the ambient medium
and the accretion flow.  If, however, accretion onto the black hole
proceeds via a Bondi or ADAF flow, a hard X-ray power law should be
detectable in deeper {\it Chandra} observations. The flux in this
power law comes from gas at $\ll R_A$ and should be $\sim 1/3$ of the
soft thermal flux from gas at $\sim R_A$ (Fig. 2a; middle curve).

It is important to stress that {\it Chandra} observations of the
Galactic Center directly determine the density and temperature of gas
at $1.5'' \approx R_A$ ($\approx 100$ cm$^{-3}$ and $\approx 2$ keV,
respectively) and at $10'' \approx 7 R_A$ ($\approx 20$ cm$^{-3}$ and
$\approx 1$ keV, respectively).  These boundary conditions strongly
constrain Bondi accretion models and it appears difficult to avoid the
hard X-ray power law seen in Figure 2 (if Bondi models are correct).

One caveat is that the models presented here assume that all of the
inflowing gas is subsonic at large radii ($\gsim 1''$).  Since Sgr A*
is believed to be fed by stellar winds with velocities $\sim 300-1000$
km s$^{-1}$ (Najarro et al. 1997), I have effectively assumed that
most of the stellar winds have shocked outside $\approx 1''$.  For the
X-ray emitting gas observed by {\it Chandra} this is probably
reasonable since the observed temperature of several keV is comparable
to that expected from the shocked stellar winds.  There could,
however, be a component of cold inflowing gas that is not accounted
for in the models presented here.

Advection-dominated accretion flow models require an additional
boundary condition on top of the density and temperature needed in
Bondi models, namely the rotation rate of the gas at $\approx R_A$.
It would be interesting to explore the predicted spectra from ADAF
models as a function of this rotation rate (which is rather
uncertain).  On physical grounds ADAF models should be very similar to
the Bondi models shown here as long as the viscous time at $\sim R_A$
is shorter than the cooling time; this is easily satisfied for the
Galactic Center provided the dimensionless viscosity $\alpha$ is
$\gsim 10^{-3}$.  My preliminary calculations of ADAF models accreting
from an ambient medium support this conclusion.

Although the presence of an underlying hard X-ray power law at roughly
the level predicted here appears relatively robust within the context
of ADAF and Bondi models, it does depend sensitively on the structure
of the accreting gas at $R \lsim R_A$. Moreover, recent theoretical
work suggests that that the dynamics of radiatively inefficient
accretion flows may be quite different from that predicted by Bondi
and ADAF models (e.g., Blandford \& Begelman 1999; Stone et al. 1999;
Igumenshchev \& Abramowicz 2000; Narayan et al. 2000; Quataert \&
Gruzinov 2000; Igumenshchev \& Narayan 2002; Hawley \& Balbus 2002).
For a parameterized density profile of the form $\rho \propto R^{-3/2
+ p}$ ($0 < p < 1$), large values of $p \sim 1/2-1$ are favored over
the Bondi/ADAF value of $p = 0$.

It is unclear whether these modifications to the structure of the
accretion flow occur as far out as $\sim R_A$ or whether they are
confined to regions closer to the black hole.  For example, most of
the physics that leads to significant deviations from ADAF/Bondi
models requires dynamically significant angular momentum (see,
however, Igumenshchev \& Narayan 2002).  It is unclear whether this is
appropriate near $\sim R_A$, where most of the detectable
bremsstrahlung emission originates.  Thus, even if the flow structure
is very different in the vicinity of the black hole, Bondi models may
be applicable near $\sim R_A$.\footnote{This is a subtle issue
because, even if the angular momentum barrier formally lies at a
radius $\ll R_A$, angular momentum can still influence the flow
structure out to $\sim R_A$.  This is because the flow is causally
connected from near the horizon out to large radii.}

If the flow structure is significantly modified in the vicinity of
$R_A$, one does not expect to see a hard X-ray power law analogous to
the Bondi/ADAF results in Figure 2. The reason is that the
bremsstrahlung spectrum from a $\rho \propto R^{-3/2 + p}$ density
profile is $\nu L_\nu \propto \nu^{1/2 - 2p}$ (Quataert \& Narayan
1999). For $p = 0$ (Bondi/ADAF), a hard X-ray power law is present
(Fig. 2) while for $p \sim 1/2-1$, the spectrum is soft and so there
should be very little hard X-ray emission. Deep {\it Chandra}
observations of the quiescent emission from the Galactic Center can
therefore shed important light on the structure of the accretion flow
onto the central black hole.  Interpreting the data may be complicated
if, as is plausible, SSC and/or synchrotron emission contribute to the
quiescent flux at some level.  The bremsstrahlung contribution can be
isolated by focusing on the least variable segments of the observed
data.  The strength of thermal X-ray lines also provides a constraint
on synchrotron or SSC contributions to the quiescent emission (Narayan
\& Raymond 1999).

The constraints on an underlying hard X-ray power law can be
significantly tightened if there are sufficient counts to fully
utilize {\it Chandra's} resolution and extract spectra from a $\approx
0.5''$ region around Sgr A*.  As Figure 2a shows, most of the soft
thermal emission would then be resolved out (since $R_{beam} \approx
0.3 R_A$).  It might even be possible to construct a radial surface
brightness profile of the outer parts of the accretion flow to compare
with theoretical models (e.g., Quataert \& Narayan 2000; \"Ozel \& Di
Matteo 2001).  For example, in ADAF/Bondi models, the spectrum should
harden significantly as $R_{beam}/R_A$ decreases from 1 to 0.3
(Fig. 2).

\subsection{Additional Applications}

The results in this paper are also useful for interpreting {\it
Chandra} observations of massive black holes in early type galaxies.
In such systems the central black hole may accrete the hot ambient
interstellar medium of the host galaxy (e.g., Fabian \& Rees 1995; Di
Matteo et al. 1999).  Loewenstein et al. (2001) reported the
nondetection of a central hard X-ray point source in NGC 1399, NGC
4472, and NGC 4636.  Given black hole mass estimates from the
$M_{bh}-\sigma$ relation (Gebhardt et al. 2000; Ferrarese \& Merritt
2000), $R_{beam} \approx 3, 4$, and $20 \ R_A$ for these systems.  The
ambient ISM in elliptical galaxies is typically stratified as $\rho
\propto R^{-1}$ so the results in Figure 2a can be used to estimate
the expected hard X-ray bremsstrahlung emission from a Bondi or ADAF
accretion flow.  I find that the hard X-ray flux should be $\sim 0.1,
0.1$, and $0.025$ of the soft X-ray flux in the central 1'' for NGC
1399, NGC 4472, and NGC 4636, respectively.  Deeper {\it Chandra}
observations of NGC 1399 and NGC 4472 may be able to detect emission
at this level and would provide interesting constraints on accretion
models.  One complication in interpreting such observations is that
the black hole mass is estimated from the $M_{bh}-\sigma$ relation,
not direct dynamical studies. This leads to a factor of few
uncertainty in $R_A$.  In this respect, M87 is a more promising system
since the black hole mass is dynamically determined to be $\approx 3
\pm 1 \times 10^9 M_\odot$ (e.g., Macchetto et al. 1997); $R_A$ is
thus $\approx$ 1'' as in the Galactic Center.  Unfortunately, however,
unresolved emission from the jet appears to dominate {\it Chandra}
observations of the nucleus of M87 (Wilson \& Yang 2002) and so hard
X-ray bremsstrahlung emission will be difficult to detect.

\acknowledgements

I thank Fred Baganoff, Andrei Gruzinov, Mike Loewenstein, Ramesh
Narayan, and the referee for useful discussions and comments.

\newpage

\begin{figure}
\plotone{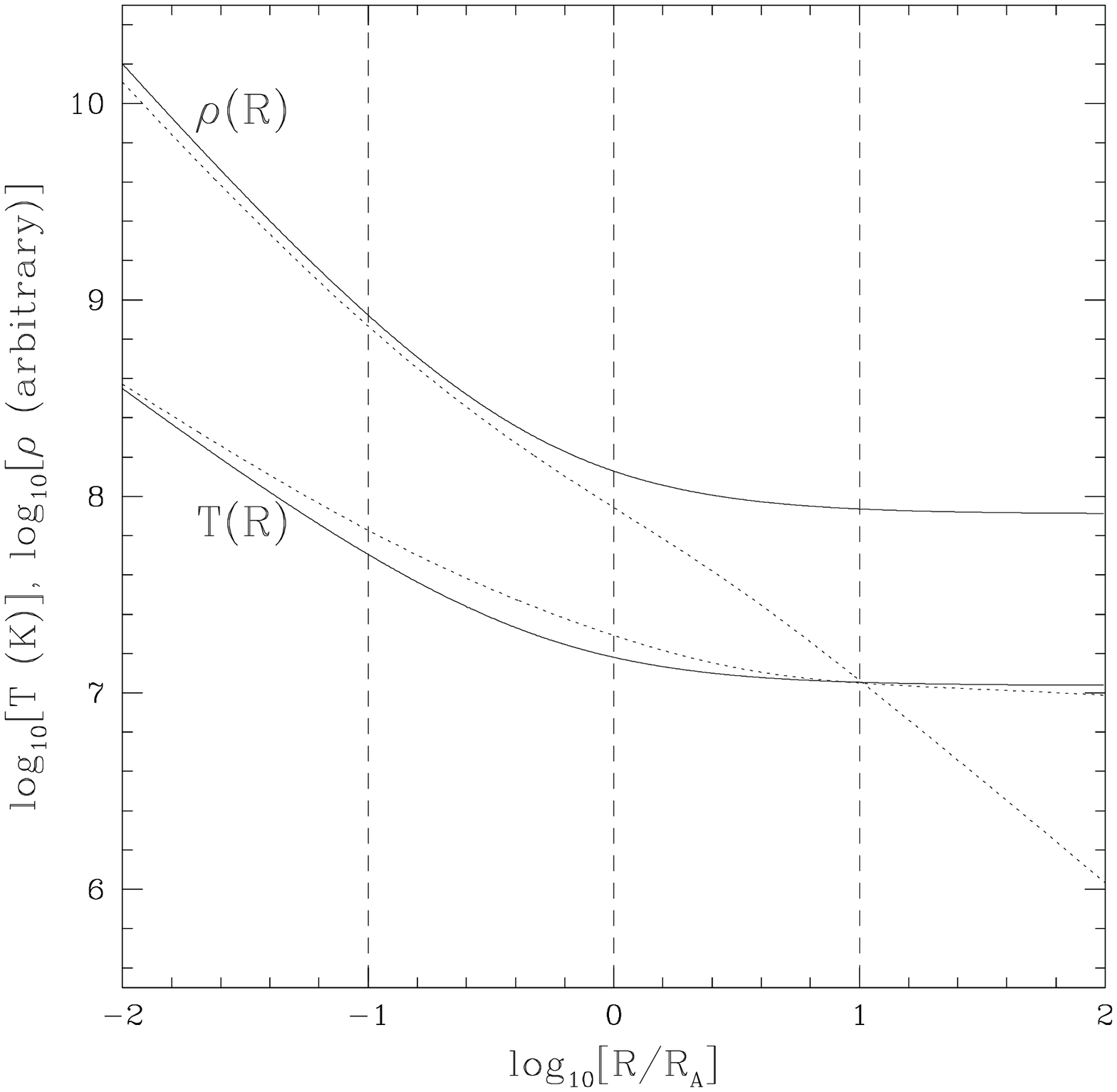}
\caption{Density and temperature profiles for spherical Bondi
accretion.  The solid lines are for accretion from a uniform ambient
medium while the dotted lines are for accretion from a stratified
medium in which the ambient density decreases with radius as $\rho
\propto R^{-1}$.  The vertical dashed lines help indicate the part of
the accretion flow probed by a telescope with $R_{beam}/R_A = 0.1, 1,$
and $10$ (from left to right).  {\it Chandra} has $R_{beam} \approx R_A$
for the Galactic Center and $R_{beam} \approx 1-10 R_A$ for a number
of elliptical galaxies in nearby X-ray clusters.}
\end{figure}

\begin{figure}
\plottwo{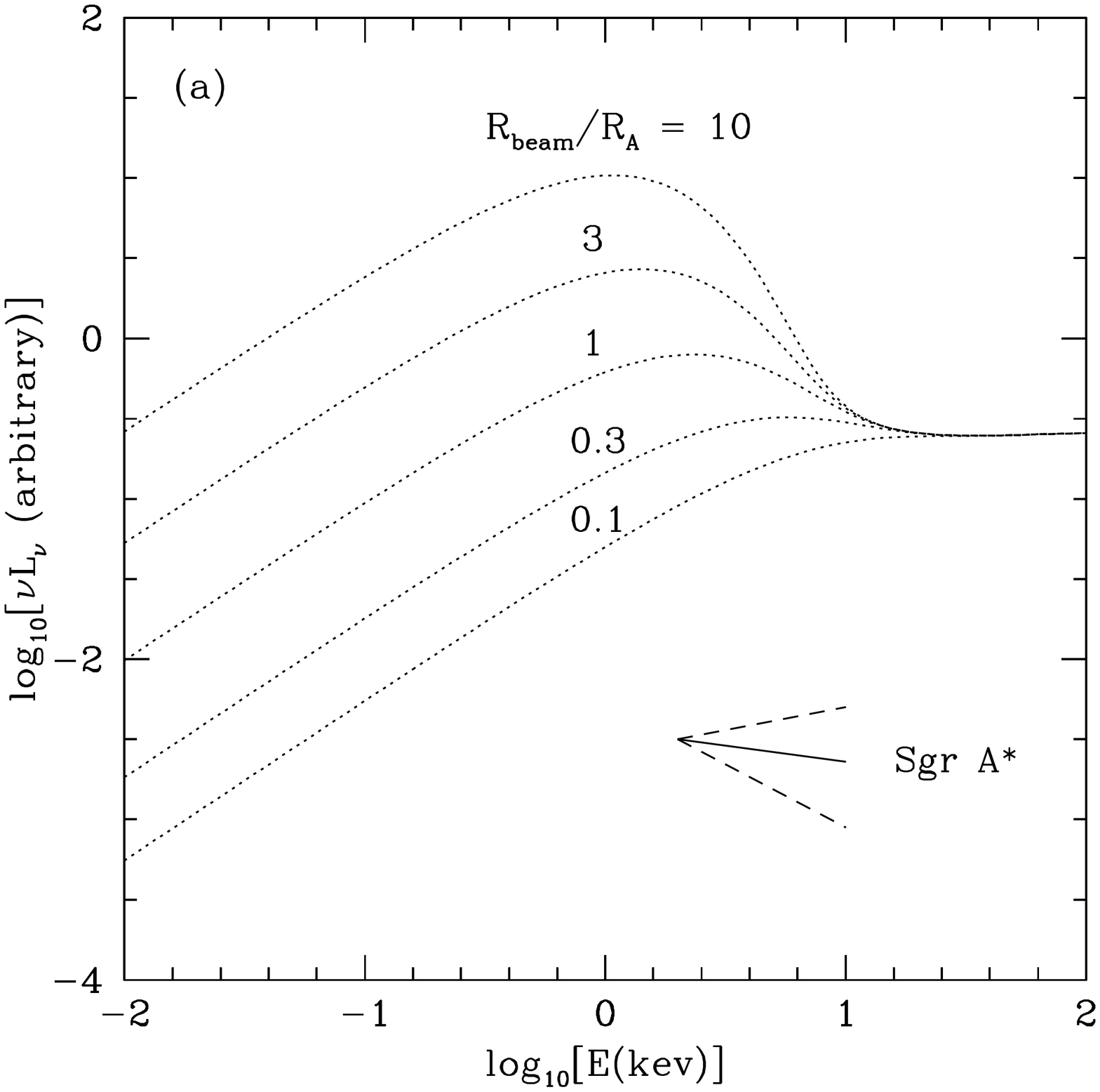}{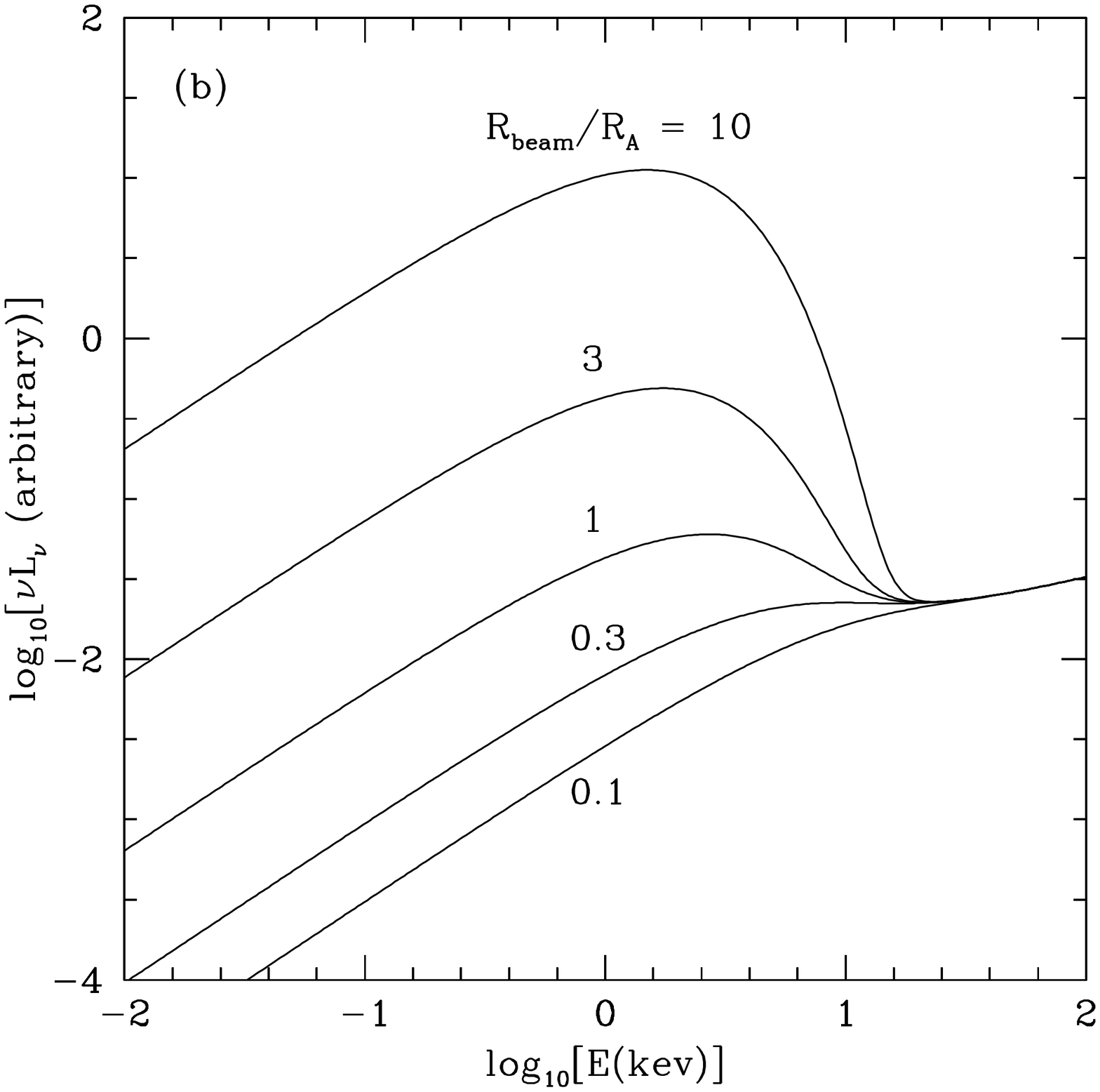}
\caption{Bremsstrahlung X-ray spectra using the density and
temperature profiles in Figure 1.  The spectra are shown for various
values of the size of the telescope beam relative to the Bondi
accretion radius.  Fig. 2a is for accretion from a radially stratified
ambient medium (dotted lines in Fig. 1) while Fig. 2b is for accretion
from a uniform ambient medium (solid lines in Fig. 1).  The best fit
spectral slope ($2-10$ keV) for quiescent emission from within 1.5''
of Sgr A* is shown by the solid line in Fig. 2a (1.5'' corresponds to
$R_{beam} \approx R_A$); the dashed lines give the $90 \%$ confidence
limits (Baganoff et al. 2001).}
\end{figure}

\end{document}